\documentclass[preprint,12pt,authoryear]{elsarticle}

\newtheorem{theorem}{Theorem}[section]

\newtheorem{lemma}[theorem]{Lemma}

\newtheorem{corollary}[theorem]{Corollary}

\newcommand{\bea}{\begin{eqnarray}}
\newcommand{\eea}{\end{eqnarray}}
\newcommand{\beax}{\begin{eqnarray*}}
\newcommand{\eeax}{\end{eqnarray*}}

\iffalse 
  
  \newcommand{\PartDoubleLine}{\setlength{\baselineskip}{2\baselineskip}}
\else    
  
  \newcommand{\PartDoubleLine}{\setlength{\baselineskip}{1\baselineskip}}
\fi
\newcommand{\BeginTableTitle}{\begin{quotation}\noindent\PartDoubleLine}
\newcommand{\EndTableTitle}{\\\end{quotation}}
\newproof{pf}{Proof}
\def\drow{\stackrel{d}{\longrightarrow}}
\def\prow{\stackrel{\textstyle p}{\longrightarrow}}

\usepackage{amssymb}

\usepackage{setspace}

\usepackage{booktabs}

\usepackage{appendix}

\journal{ }

\begin{document}

\begin{frontmatter}



\title{Weighted empirical likelihood for quantile regression with nonignorable missing covariates}


\author{Xiaohui Yuan}
\ead{yuanxh@ccut.edu.cn}
\author{Xiaogang Dong}
\ead{dongxiaogang@ccut.edu.cn}
\cortext[cor1]{Corresponding author.}

\address{School of Basic Science, Changchun University of Technology, Changchun 130012, China}

\begin{abstract}
In this paper, we propose an empirical likelihood-based weighted estimator of regression parameter in quantile regression model with nonignorable missing covariates. The proposed estimator is computationally simple and achieves semiparametric efficiency if the probability of missingness on the fully observed variables is correctly specified. The efficiency gain of the proposed estimator over the complete-case-analysis estimator is quantified theoretically and illustrated via simulation and a real data application.

\end{abstract}


\end{frontmatter}



\section{Introduction}

Quantile regression, as introduced by Koenker and Bassett (1978), is robust against outliers and can describe the entire conditional distribution of the response variable given the covariates. Due to these advantages, quantile regression became appealing in econometrics, statistics, and biostatistics. The book by Koenker (2005) contains a comprehensive account of overview and discussions in quantile regression.

Let $Y$ denote the outcome variable,  $Z$ be  a vector of covariates which is always observed, and $X$ be a vector of covariates which may not be observed for all subjects.
The quantile regression model assumes that the $\tau$-th conditional quantile of $Y$ given $X$ and $Z$:
\begin{eqnarray}
Q_{\tau}(Y|X,Z,\beta^*)=\beta_0^*+X^T\beta_1^*+Z^T\beta_2^*=W^T\beta^*,
\label{model}
\end{eqnarray}
where $W=(1,X^T,Z^T)^T$ and $\beta^*=(\beta_0^*,\beta_1^{*T},\beta_2^{*T})^T$ is interior to parameter space $\Theta$, $\Theta$ is a compact subset of $R^p$.   We are interested in the inference about $\beta^*$ based on a random sample of incomplete  data
\begin{eqnarray*}
 (Y_i,X_i^T,Z_i^T,\delta_i),\ \ i=1,\cdots,n,
\label{data}
\end{eqnarray*}
where all the $Z_i$'s and $Y_i$'s are observed, and $\delta_i = 0$ if $X_i$ is missing, otherwise $\delta_i = 1$.

The most commonly used method for handling missing covariate
data is the complete-case analysis (CCA), with only the remaining complete data used to perform a
regression-based or likelihood-based analysis. The CCA esitmator of $\beta^*$ is given by
\begin{eqnarray}
\hat{\beta}_C=\arg\min_{\beta\in\Theta}\frac{1}{n}\sum_{i=1}^n\delta_i\rho_{\tau}(Y_i-W_{i}^T\beta),
\label{mfunc}
\end{eqnarray}
where $\rho_{\tau}(u)=u\{\tau-I(u<0)\}$ is the quantile loss function and $I(\cdot)$ is the indicator function.

In statistic literature, there are three missing data categories  (Little and Rubin, 2002).  The first case is
missing completely at random (MCAR), i.e., data missing mechanism is independent of any
observable or unobservable quantities.
The second case is  missing at random (MAR), i.e., data missing mechanism
depends on the observed variables. The third case is not missing at random (NMAR) or nonignorable, i.e., data
missing mechanism depends on their own values.

When $X_i$'s are not MCAR, the CCA estimator can be biased. Consistent and efficient estimators have been proposed in the statistical literature for the quantile regression model when the covariates data are MAR. See for example, Wei et al. (2012)  developed an iterative imputation procedure for estimating the conditional quantile in the presence of missing covariates. Sherwood et al. (2013) proposed an inverse probability weighted (IPW) approach to correct for the bias from longitudinal dropouts. Chen et al. (2015) examined the problem of estimation in a quantile regression model and  developed three nonparametric methods when observations are missing at random under independent and nonidentically distributed errors.
Liu and Yuan (2016) proposed a weighted quantile regression model with weights chosen by empirical likelihood. This approach efficiently incorporates the incomplete data into the data analysis by combining the complete data unbiased estimating equations and incomplete data unbiased estimating equations. However, it may not be an easy task to extend these methods to deal with NMAR missing data mechanisms, because these methods are biased under the NMAR assumption.

NMAR is the most difficult problem in the missing data literature. Following Little and Zhang (2011) and Bartlett et al. (2014), we make the following ``not missing at random'' (NMAR) assumption:
\begin{eqnarray}
Y\perp \delta |X,Z.
\label{miss}
\end{eqnarray}
The NMAR assumption (\ref{miss}) implies that, missingness in a covariate depends on the value of that covariate, but is conditionally independent of outcome. The CCA estimator is valid but inefficient under the assumption (\ref{miss}) because it fails to draw on the observed information contained in the incomplete cases.

In the context of mean regression model, Bartlett et al. (2014) proposed an augmented CCA estimator to improve upon the efficiency of CCA estimator by modeling an additional model for the probability of missingness on the fully observed variables, i.e. $P(\delta=1|Y,Z)$. The estimating function used in Bartlett et al. (2014) utilizes all the observed data by drawing on the information available from both complete and incomplete cases and thus improves upon the efficiency of CCA estimator. Note that under NMAR assumption (\ref{miss}), $P(\delta=1|Y,X,Z)=P(\delta=1|X,Z)$, whose feasible estimators are not available, since the observations of $X$ are missing on some subjects. Thanks to the NMAR assumption (\ref{miss}), there is no need to estimate $P(\delta=1|X,Z)$ under the assumption (\ref{miss}). Recently, Xie and Zhang (2017) proposed an empirical likelihood approach for estimating the regression parameters in mean regression model with missing covariates under NMAR assumption (\ref{miss}). They showed that the empirical likelihood estimator can improve estimation efficiency if $P(\delta=1|Y,Z)$ is correctly specified.

In this paper, we put forward an empirical likelihood-based weighted (ELW) estimator for estimating quantile regression model with nonignorable missing covariates under NMAR assumption (\ref{miss}). To fully utilize the information contained in the incomplete data, we incorporate the unbiased estimating equations of incomplete observations into empirical likelihood and obtain the empirical likelihood-based weights to adjust the CCA estimator defined in (\ref{mfunc}). The proposed ELW estimator is computationally simple as the CCA estimator and achieves semiparametric efficiency if $P(\delta=1|Y,Z)$ is correctly specified.

Empirical likelihood is an effective approach to improving efficiency. For a comprehensive review of the empirical likelihood
method, one can refer to  Qin and Lawless (1994), Owen (2001), Lopez et al. (2009)   among others.
For applications of empirical likelihood in missing-data problems, one can refer to Wang and Rao (2002), Qin et al. (2009),  Liu and Yuan (2012), Liu et al. (2013),  Zhong and Qin (2017) among others.

The rest of this paper is organized as follows. In section 2, we introduce the empirical likelihood-based weighted estimator for quantile regression model.
In section 3, we show that the ELW estimator is asymptotically equivalent to the profile empirical likelihood estimator and thus achieves semiparametric efficiency.
Numerical studies are reported in sections 4-5.  Proofs of the main theorems  needed are given in the Appendix.

\section{The empirical likelihood-based weighted estimation}
\noindent
In this section, we propose the ELW estimator of $\beta^*$ under the assumption (\ref{miss}).
Under the assumption (\ref{miss}), we only need to estimate the probability of $X$ being observed given $Y$ and $Z$, i.e. $P(\delta=1|Y,Z)$. Following Bartlett et al. (2014) and Xie and Zhang (2017), we assume that $P(\delta=1|Y,Z)$ is described by the probability model:
 \begin{eqnarray}\label{dyz}
P(\delta=1|Y,Z)=\pi(Y,Z,\gamma^*),
\end{eqnarray}
where $\gamma^*$ is a $q \times 1$ unknown vector parameter. It is natural to estimate $\gamma^*$ by the binomial likelihood estimator $\hat{\gamma}$ which maximizes the binomial log-likelihood
\begin{eqnarray*}
L_B(\gamma)=\sum_{i=1}^n[\delta_i\log\{\pi(Y_i,Z_i,\gamma)\}+(1-\delta_i)\log\{1-\pi(Y_i,Z_i,\gamma)\}].
\end{eqnarray*}
Let $m(Y_i,Z_i, \beta,\alpha)$ be a working model of $E\{\delta_i\phi(X_i,Z_i,Y_i,\beta)|Z_i,Y_i\}$ with $\phi(X_i,Z_i,Y_i,\beta)=W_i\{I(Y_i-W_{i}^T\beta<0)-\tau\}$. In the following, we proposed the ELW estimator of $\beta^*$.
Define
\begin{eqnarray*}
&&U_{B}(\delta_i,Z_i,Y_i,\gamma)=\frac{ \delta_i- \pi(Y_i,Z_i,\gamma)}{\pi(Y_i,Z_i,\gamma)\{1-\pi(Y_i,Z_i,\gamma)\} }\frac{\partial \pi(Y_i,Z_i,\gamma)}{\partial \gamma},\\
&&g_1(\delta_i,X_i,Z_i,Y_i,\theta)= [\delta_i- \pi(Y_i,Z_i,\gamma)]m(Z_i,Y_i,\beta,\alpha ),\\
&&g(\delta_i,X_i,Z_i,Y_i,\theta)=\left(\begin{array}{c}
        g_1(\delta_i,X_i,Z_i,Y_i,\theta) \\
         U_{B}(\delta_i,Z_i,Y_i,\gamma)
       \end{array}\right).
\end{eqnarray*}
Let $p_i$ represent the probability weight allocated to $ g(\delta_i,X_i,Z_i,Y_i,\hat{\theta})$, where $\hat{\theta}=(\hat{\alpha}^T,\hat{\beta}_C^T, \hat{\gamma}^T)^T$ and $\hat{\alpha}$   is a consistent
estimator for some $\alpha^*$. If $\pi(y,z,\gamma)$ is correctly specified, one can show that $E\{g(\delta_i,X_i,Z_i,Y_i,\theta^*)\}=0$, where $\theta^*=(\alpha^{*T},\beta^{*T}, \hat{\gamma}^{*T})^T$. Then, we maximize the empirical likelihood function $\prod_{i=1}^np_i$
subject to the constraints:
\begin{eqnarray*}
p_i\geq0,\ \ \ \sum_{i=1}^np_{i}=1,\ \ \
\sum_{i=1}^np_{i}g(\delta_i,X_i,Z_i,Y_i,\hat{\theta})=0.
\end{eqnarray*}
By using the Lagrange multiplier method,  we get
\begin{eqnarray*}
\hat{p}_i=\frac{1}{n}\frac{1}{1+\hat{\lambda}^Tg(\delta_i,X_i,Z_i,Y_i,\hat{\theta})},
\label{pi}
\end{eqnarray*}
where $\hat{\lambda}$ is the Lagrange multiplier that satisfies
\begin{eqnarray*}
\frac{1}{n}\sum_{i=1}^n\frac{g(\delta_i,X_i,Z_i,Y_i,\hat{\theta})}
{1+\hat{\lambda}^Tg(\delta_i,X_i,Z_i,Y_i,\hat{\theta})}=0.
\label{lam}
\end{eqnarray*}
The ELW estimator of $\beta^*$ is given by
\begin{eqnarray}
\hat{\beta}_{ELW}=\arg\min_{\beta\in\Theta}\sum_{i=1}^n \hat{p}_i\delta_i \rho_{\tau}(Y_i-W_{i}^T\beta).
\label{elw}
\end{eqnarray}
Define
\begin{eqnarray}
\lambda(\theta)=\arg\max_{\lambda}\sum_{i=1}^n\log\{1+\lambda^T g(\delta_i,X_i,Z_i,Y_i,\theta)\}.
\label{lamconvexp}
\end{eqnarray}
From (\ref{lam}), it is easily seen $\hat{\lambda}=\lambda(\hat{\theta})$. For fixed $\theta=\hat{\theta}$, solving (\ref{lamconvexp}) is a well-behaved optimization problem since the objective function is globally concave and can be solved by a simple Newton-Raphson numerical
procedure.

Let $F_{i}(\cdot)$ and $f_{i}(\cdot)$ denote respectively the conditional distribution and density functions of $Y_i$ given $(X_i,Z_i)$.  Denote
\begin{eqnarray*}
  F_\beta &=& E\left\{ \delta_if_{i}(0) W_{i}W_{i}^T\right\},\\
  S_\phi&=&E\left\{\delta_i\phi(X_i,Z_i,Y_i,\beta^*)\phi^T(X_i,Z_i,Y_i,\beta^*)\right\},\\
D_1&=&E\left\{[\delta_i- \pi(Y_i,Z_i,\gamma^*)]^2 m(Z_i,Y_i,\beta^*,\alpha^* )m^T(Z_i,Y_i,\beta^*,\alpha^* )  \right\},\\
D_2&=&E\left\{[\delta_i- \pi(Y_i,Z_i,\gamma^*)]m(Z_i,Y_i,\beta^*,\alpha^* )U_{B}^T(\delta_i,Z_i,Y_i,\gamma^*)\right\},\\
D_3&=&E\left\{ \delta_i[\delta_i- \pi(Y_i,Z_i,\gamma^*)] \phi(X_i,Z_i,Y_i,\beta^*) m^T(Z_i,Y_i,\beta^*,\alpha^* )   \right\},\\
D_4&=&E\left\{ \delta_i  \phi(X_i,Z_i,Y_i,\beta^*)U_{B}^T(\delta_i,Z_i,Y_i,\gamma^*)   \right\},\\
S_B&=&E\left\{U_{B}(\delta_i,Z_i,Y_i,\gamma^*)  U_{B}^T(\delta_i,Z_i,Y_i,\gamma^*)   \right\}.
\end{eqnarray*}The following regularity conditions help us in doing asymptotic analysis:
\begin{enumerate}

\item[C1] The $\tau$-th conditional quantile of $Y_i$ given $W_i$ is $Q_{\tau}(Y_i|W_i,\beta^*)=W_i^T\beta^*$ and $W_i$ has a bounded support.
\item[C2] $Y\perp \delta |X,Z$.
\item[C3] $F_\beta$, $S_\phi$, $S_B$  are positive definite.
\item[C4] $F_{i}(\cdot)$ is absolutely continuous and $f_{i}(\cdot)$ is uniformly bounded away from 0 and $\infty$ at 0.
\item[C5] (a) $P(\delta=1|Y,Z)=\pi(Y,Z,\gamma^*)$  (b)  $\inf_{(Y,Z)}\pi(Y,Z,\gamma^*) \geq c_0$ for some $c_0>0.$ (c) For all $(Y_{i},Z_{i})$, $\pi(Y_{i},Z_{i},\gamma)$
admits all third partial derivatives $\frac{\partial^3\pi(Y_{i},Z_{i},\gamma)}{\partial\gamma_k\partial\gamma_l\partial\gamma_m}$
for all $\gamma$ in a neighborhood of the true value $\gamma^*$,
$\biggr\|\frac{\partial^3\pi(Y_{i},Z_{i},\gamma)}{\partial\gamma_k\partial\gamma_l\partial\gamma_m}\biggr\|$ and  $\|\partial\pi(Y_i,Z_i,\gamma)/\partial\gamma\|^2$
are bounded by an integrable function for all $\gamma$ in this neighborhood.
\item[C6] For all $(Y_{i},Z_{i})$, $m(Y_i,Z_i,\beta,\alpha)$
admits all second partial derivatives $\frac{ \partial^2 m(Y_i,Z_i,\beta,\alpha)}{\partial \beta_i \partial \beta_j}$ and $ \frac{\partial^2 m(Y_i,Z_i,\beta,\alpha)}{\partial \alpha_i \partial \alpha_j}$
for all $\beta$ and $\alpha$ in a neighborhood of $(\beta^{*T},\alpha^{*T})^T$. $\|m(Y_i,Z_i,\beta,\alpha)\|^2 $, $\|\frac{\partial^2 m(Y_i,Z_i,\beta,\alpha)}{\partial \beta_i \partial \beta_j}\|$ and $\|\frac{\partial^2 m(Y_i,Z_i,\beta,\alpha)}{\partial \alpha_i \partial \alpha_j}\|$  are bounded by an integrable function for all $\beta$ and $\alpha$ in this neighborhood.

\end{enumerate}

The asymptotic distribution of $\hat{\beta}_{C}$ is given by the following theorem.
\begin{theorem}\label{ccatheorem}

Under conditions C1-C4, $n^{1/2}(\hat{\beta}_{C}-\beta^*)\drow N(0,\Sigma_{C})$ as $n\rightarrow\infty$, where $\Sigma_{C}=F_\beta^{-1} S_\phi F_\beta^{-1}.$
\end{theorem}

The asymptotic distribution of $\hat{\beta}_{ELW}$   is given by the following theorem.

\begin{theorem}\label{elwnormtheorm}
Under conditions C1-C6,  $n^{1/2}(\hat{\beta}_{ELW}-\beta^*)\drow N(0,\Sigma_{ELW})$ as $n\rightarrow\infty$, where
\begin{eqnarray*}
  \Sigma_{ELW} &=& F_{\beta}^{-1}\left(S_\phi-V_1 V_2^{-1}V_1^T \right)F_{\beta}^{-1}
               =  \Sigma_{C}- F_{\beta}^{-1}V_1 V_2^{-1}V_1^T  F_{\beta}^{-1},
\end{eqnarray*}
  $V_1 = D_3-D_4S_B^{-1}D_2^T$ and $V_2 = D_1-D_2S_B^{-1}D_2^T.$
\end{theorem}

For two matrices $A$ and $B$, we write $A \leq B$ if $B - A$ is a nonnegative-definite matrix.
\begin{corollary}\label{AB}
If both $F_\beta$ and $V_2$ are positive definite, we
have $\Sigma_{ELW}\leq \Sigma_{C}$, and the equality holds if and only if
$V_1=0$.
\end{corollary}

Corollary \ref{AB} reveals that  $\hat{\beta}_{ELW}$ is at least as efficient as  $\hat{\beta}_{C}$ for any
working regression function $m(Y_i,Z_i,\beta,\alpha)$, whether or not it correctly identifies the optimal regression
function $E\{\phi(X_i,Z_i,Y_i,\beta)|Z_i,Y_i,\delta_i=1\}$.

Although $\hat{\beta}_{ELW}$ can be obtained easily, it is  difficult to estimate the limiting covariance matrix analytically. We apply the resampling method in Liu and Yuan (2016) to the inference about $\beta^*$.

\section{Simulation studies}
\noindent In this section, we investigate the performance of the proposed estimator $\hat{\beta}_{ELW}$  and several other estimators based on Monte-Carlo simulations.

The simulated data are generated by the procedure of Bartlett et al. (2014), in which the non-missing indicator $\delta$ is distributed with $P(\delta=1)=0.5$, and $(X,Z,Y)$ is generated from a trivariate normal distribution conditional on $\delta$:  $$(X,Z,Y)^T|\delta   \thicksim N( (\delta, 0, \eta \delta)^T, \Psi),$$
where $\Psi=(\sigma_{a b})$, $a,b=x,z,y$, $\eta= (\sigma_{xy}\sigma_{zz}-\sigma_{xz}\sigma_{zy})\upsilon_1$ and $\upsilon_1=(\sigma_{xx}\sigma_{zz}-\sigma_{xz}^2)^{-1} $.

It is easy to verify that the assumption $\delta\bot Y|(X,Z)$ is satisfied in this setup.  Conditional on $Z$ and $Y$, the probability of $P(\delta=1|Z,Y)$ is a logistic regression with
$$P(\delta=1|Z,Y)=\frac{\exp(\gamma_0+\gamma_1Z+\gamma_2Y)}{1+\exp(\gamma_0+\gamma_1Z+\gamma_2Y)}$$
where $\gamma_0=- {0.5\eta^2\sigma_{zz}}\upsilon_2,
\gamma_1=- {\eta\sigma_{zy}}\upsilon_2,
\gamma_2= {\eta\sigma_{zz}}\upsilon_2 $ and $\upsilon_2=(\sigma_{zz}\sigma_{yy}- \sigma_{zy}^2)^{-1} $.
  The conditional quantile model of interest is specified as $$Q_\tau(Y|X,Z)= \beta_0+ \beta_1X+\beta_2Z, $$with $\beta_0=\Phi^{-1}(\tau,\sigma^2),$ $ \beta_1=(\sigma_{xy}\sigma_{zz}-\sigma_{xz}\sigma_{zy})\upsilon_1$,
$\beta_2=(\sigma_{zy}\sigma_{xx}-\sigma_{xz}\sigma_{xy})\upsilon_1$, $\sigma^2=\sigma_{yy}- ( \sigma_{xz}^2\sigma_{zz}-2\sigma_{xz}^2\sigma_{zy} + \sigma_{zy}^2\sigma_{xx})\upsilon_1$.

We set  $\sigma_{xx}=\sigma_{yy}=\sigma_{zz}=1$, $\sigma_{xz}=\sigma_{xy}=\sigma_{zy}=0.5$ and generate 1000 Monte Carlo data sets of sample sizes $n=100$ and $300$.
Five  estimators are considered:
\begin{enumerate}
  \item[1.] $\hat{\beta}_{ideal}$: the  quantile regression estimator with the full observations. This is the ideal case, but it is not feasible in practice. Nevertheless, we used it as a benchmark for comparison;
  \item[2.] $\hat{\beta}_C$: the CCA estimator defined in equation (\ref{mfunc});
  \item[3.] $\hat{\beta}_{IPWMAR}$: the IPW estimator assuming MAR,  introduced in Sherwood et al. (2013);
  \item[4.] $\hat{\beta}_{ELWMAR}$: the ELW estimator assuming MAR,  proposed by Liu and Yuan (2016);
  \item[5.] $\hat{\beta}_{ELW}$: the ELW estimator defined in equation (\ref{elw}).
\end{enumerate}

The empirical bias and the root-mean-squared errors (RMSEs) of the proposed estimators  with sample sizes of 100 and 300 are reported in Table 1.
 The results can be summarized as follows: the CCA estimator $\hat{\beta}_C$  and the ELW estimator $\hat{\beta}_{ELW}$ are unbiased as expected. While  $\hat{\beta}_{IPWMAR}$ and  $\hat{\beta}_{ELWMAR}$ for $\beta_0$  are clearly biased.
$\hat{\beta}_{ELW}$  performs better than $\hat{\beta}_C$  in terms
of RMSE  in most cases, which agrees with our theory.  $\hat{\beta}_{C}$ and $\hat{\beta}_{ELW}$ are  improved in terms of RMSE as the sample size $n$ goes up from 100 to 300.

\section{Data analysis}

In this section, we  apply the proposed method to the data on alcohol consumption, age, body mass index and systolic blood pressure from the 2013-2014 NHANES.
We model the population quantile of SBP (systolic blood pressure)  as a function of the  following four covariates: BMI (body mass index),
 Alcohol (log\{alcohol consumption per day$+1$\}), Age ($\{\mbox{age}-50\}/10$) and  Age$^2$ ($\{\mbox{age}-50\}^2/100$).

In our analysis, there are 7104 observations in the data set, where the dependent variable SBP and the covariates BMI and Age have complete data, the covariate Alcohol are  missing 53.29\%. It is a priori plausible that missingness in Alcohol is primarily dependent on the value of itself (i.e. MNAR), and that missingness in Alcohol is independent of SBP conditional on Alcohol, BMI, Age, and Age$^2$. Consequently, CCA is expected to give valid inferences, while the MAR assumption likely does not hold.

For $i=1,\cdots,n=7104$, let $Y_i$ denote the $i$th observation of $Y=$SBP, $Z_i$
denote the $i$th observation of $Z$=(BMI, Age, Age$^2$)$^T$ and $X_i$ denote the $i$th observation of $X=$Alcohol. Then, we consider the following model for the $\tau$th conditional
quantile of $Y_i$ given $W_i=(1,X_i,Z_i^T)^T$:
$$Q_{\tau}(Y_i|X_i,Z_i,\beta)=\beta_0+X_i\beta_1+Z_i^T\beta_2, i=1,\cdots,n, $$
where $\beta=(\beta_0,\beta_1,\beta_2^{T})^T$ and $\beta_2=(\beta_{21},\beta_{22},\beta_{23})^T$. We consider two estimators $\hat{\beta}_C$ and $\hat{\beta}_{ELW}$. For the ELW method, the probability of  whether the Alcohol is observed is modeled by $\pi(Y,Z,\gamma)=\{1+\exp(-\gamma_0-Y\gamma_1-Z^T\gamma_2)\}^{-1}$.

In Figure 1, we plot the estimated regression coefficients, $\hat{\beta}_C$ and $\hat{\beta}_{ELW}$ for $\beta_1$, $\beta_{21}$, $\beta_{22}$ and $\beta_{23}$, at quantile levels $\tau=0.1,0.2,\cdots,0.9$. We see that the CCA and ELW methods produce similar estimated regression coefficients. In Figure 2, we plot the standard errors of $\hat{\beta}_C$ and $\hat{\beta}_{ELW}$ for $\beta_1$, $\beta_{21}$, $\beta_{22}$ and $\beta_{23}$ at various quantile levels. The standard error of $\hat{\beta}_{ELW}$ is smaller than that of $\hat{\beta}_C$ in most cases.

\section{ Conclusions}

\noindent In this paper, we develop weighted empirical likelihood approach for estimating the conditional quantile functions in linear models with nonignorable missing covariates.
By incorporating the unbiased estimating equations of incomplete data into empirical likelihood, the ELW estimator can achieve semiparametric efficiency if the probability of missingness is correctly specified. We will extend the proposed methods to other regression models, which will be investigated in the future work.

\section*{Acknowledgements}

  Xiaohui Yuan was partly supported by the NSFC (No.11401048, 11671054,11701043). Xiaogang Dong was partly supported by the NSFC (No. 11571051).



\section{Appendix}

In the section, we list a preliminary lemma which has been used in the proofs of the main results in section 2.

\begin{lemma}\label{lambdapro}
Under conditions C1-C5, we have
\begin{eqnarray*}
\hat{\lambda}=\lambda(\hat{\theta})= n^{-1}S_g^{-1}\left[U_g(\theta^*) +  G_\gamma S_B^{-1}  U(\gamma^*)\right]+o_p(n^{-1/2}),
\end{eqnarray*}
where $\lambda(\theta)$ is defined in (\ref{lamconvexp}).
\end{lemma}
\textbf{\emph{ The proof of Lemma \ref{lambdapro} }}
By Lemma A.2 in  Liu and Yuan (2016), we have
\begin{eqnarray*}\label{lamexp}
\hat{\lambda}&=&\left\{\frac{1}{n}\sum_{i=1}^n g(\delta_i,X_i,Z_i,Y_i,\hat{\theta})g^T(\delta_i,X_i,Z_i,Y_i,\hat{\theta})\right\}^{-1} n^{-1}U_g(\hat{\theta})+o_p(n^{-1/2}),\nonumber\\
\end{eqnarray*}
where $U_g(\theta)=\sum_{i=1}^ng(\delta_i,X_i,Z_i,Y_i,\theta)$. By a Taylor expansion,
\begin{eqnarray}\label{Ugg}
&&n^{-1}U_g(\hat{\theta})
\nonumber\\
&=&  n^{-1}U_g(\theta^*)+n^{-1}\frac{\partial U_g(\tilde{\theta})}{\partial \alpha^T} ( \hat{\alpha}-\alpha^*)+n^{-1}\frac{\partial U_g(\tilde{\theta})}{\partial \beta^T} ( \hat{\beta}_C-\beta^*)+  n^{-1}\frac{\partial U_g(\tilde{\theta})}{\partial \gamma^T} ( \hat{\gamma}-\gamma^*),
\nonumber\\
\end{eqnarray}
where $\tilde{\theta}$ is a point on the segment connecting $\hat{\theta}$ and $\theta^*$.
 By the  law of large numbers, we have
 \begin{eqnarray*}
   && \frac{1}{n}\sum_{i=1}^n g(\delta_i,X_i,Z_i,Y_i,\hat{\theta})g^T(\delta_i,X_i,Z_i,Y_i,\hat{\theta})\prow\left(\begin{array}{cc}
                                                                                              D_1& D_2 \\
                                                                                              D_2^T & S_B                                                                                          \end{array}\right)=S_g,\\
   && n^{-1}\frac{\partial U_g(\tilde{\theta})}{\partial \gamma^T}\prow E\left\{\frac{\partial g(\delta_i,X_i,Z_i,Y_i ,\theta^* )}{\partial \gamma^T}\right\}=\left(\begin{array}{c} -D_2 \\ -S_B \end{array} \right)=G_\gamma,\\
   &&n^{-1}\frac{\partial U_g(\tilde{\theta})}{\partial \alpha^T}\prow 0, \ \ n^{-1}\frac{\partial U_g(\tilde{\theta})}{\partial \beta^T}\prow 0.
 \end{eqnarray*}
By the asymptotic properties of maximum likelihood estimate, we have
\begin{equation}\label{hatGam}
   \hat{\gamma}-\gamma^* = n^{-1}S_B^{-1}U(\gamma^*)+o_p(n^{-1/2}),
\end{equation}
  where $U(\gamma^*)=\sum_{i=1}^nU_{B}(\delta_i,Z_i,Y_i,\gamma^*)$. Thus
  by (\ref{Ugg}) and (\ref{hatGam}),
 \begin{eqnarray*}
\hat{\lambda}&=& S_g^{-1}\left\{ n^{-1}U_g(\theta^*)+ n^{-1}\frac{\partial U_g(\tilde{\theta})}{\partial \gamma^T} ( \hat{\gamma}-\gamma^*)\right\}+o_p(n^{-1/2})  \\
&=&n^{-1}S_g^{-1}\left[U_g(\theta^*) +  G_\gamma S_B^{-1}  U(\gamma^*)\right]+o_p(n^{-1/2}).
\end{eqnarray*}
The desired result follows.
 \\
  \textbf{\emph{The proof of Theorem \ref{ccatheorem} }}
The proof is similar to  the proof of Theorem 4.1 in Koenker (2005, page 120).
 \\
\textbf{\emph{ The proof of Theorem \ref{elwnormtheorm} }}
For $i=1,\cdots,n$, let
$$A_{i}(\eta)=\rho_\tau(\varepsilon_{i}-W_{i}^T\eta/\sqrt{n})-\rho_\tau(\varepsilon_{i}),$$
 where $\varepsilon_{i}=Y_{i}-W_{i}^T\beta^*$.
The function $A(\eta)=\sum_{i=1}^n n\hat{p}_i\delta_i  A_{i}(\eta)$ is convex and is minimized at $\hat{\eta}=\sqrt{n}(\hat{\beta}_{ELW}-\beta^*)$. Following Knight's identity (Knight,1998)
$$\rho(u-v)-\rho(u)=v[I(u<0)-\tau]+\int_0^v[I(u\leq s)-I(u\leq0) ]ds,$$
we can write
$A(\eta)=A_1(\eta)+A_2(\eta),$
where
\begin{eqnarray}
 A_1(\eta) &=&n^{-1/2}\eta^T  \sum_{i=1}^n n\hat{p}_i\delta_i \phi(X_i,Z_i,Y_i,\beta^*),\label{AA1}
\\
A_2(\eta)&=&\sum_{i=1}^nn\hat{p}_i\delta_i \int_0^{w_{i}^T\eta/\sqrt{n}}\{I(\varepsilon_{i}\leq s)-I(\varepsilon_{i}\leq 0)\}ds.\label{AA2}
\end{eqnarray}

We first give the asymptotic expression of (\ref{AA1}).
Applying a Taylor expansion, we get
\begin{eqnarray*}
&&\sum_{i=1}^n n\hat{p}_i\delta_i A_{1i}(\eta) \\
&=&\eta^Tn^{-1/2}\sum_{i=1}^n \delta_i\phi(X_i,Z_i,Y_i,\beta^*)\\
&&-\eta^Tn^{-1}\sum_{i=1}^n \delta_i \phi(X_i,Z_i,Y_i,\beta^*)g^T(\delta_i,X_i,Z_i,Y_i, \theta^* ) n^{1/2}\hat{\lambda}+o_p(1).
\end{eqnarray*}
By the  law of large numbers, we have
\begin{equation}\label{Fg}
 n^{-1}\sum_{i=1}^n \delta_i \phi(X_i,Z_i,Y_i,\beta^*)g^T(\delta_i,X_i,Z_i,Y_i, \theta^* )\prow F_g=\left(\begin{array}{cc} D_3 & D_4 \end{array}\right).
\end{equation}
By Lemma \ref{lambdapro},
\begin{eqnarray*}
&&\sum_{i=1}^n n\hat{p}_i\delta_i A_{1i}(\eta) \\
&=&\eta^Tn^{-1/2}\sum_{i=1}^n \delta_i\phi(X_i,Z_i,Y_i,\beta^*)-\eta^TF_g n^{1/2}\hat{\lambda}+o_p(1)\\
&=&\eta^Tn^{-1/2}\left\{  U_\phi(\theta^*)-F_gS_g^{-1} \left[U_g(\theta^*) +  G_\gamma S_B^{-1}  U(\gamma^*)\right]\right\}+o_p(1),
\end{eqnarray*}
where $U_\phi(\theta^*)=\sum_{i=1}^n \delta_i\phi(X_i,Z_i,Y_i,\beta^*)$.

Next, we give the asymptotic expression of (\ref{AA2}). A Taylor expansion reveals that
\begin{eqnarray*}
\sum_{i=1}^n n\hat{p}_i\delta_i A_{2i}(\eta)
&=&\sum_{i=1}^n \delta_i A_{2i}(\eta) -\sum_{i=1}^nA_{2i}(\eta)\delta_ig^T(\delta_i,X_i,Z_i,Y_i, \theta^*)\hat{\lambda}+o_p(1).
\end{eqnarray*}
Moreover, similar to the proof of Theorem 4.1 in  Koenker(2005), one can show that
\begin{eqnarray*}
\sum_{i=1}^n \delta_i A_{2i}(\eta)&=&\frac{1}{2}\eta^T F_\beta\eta +o_p(1).
\label{A21}
\end{eqnarray*}
Thus, we only need to show that $\sum_{i=1}^nA_{2i}(\eta)\delta_ig^T(\delta_i,X_i,Z_i,Y_i, \theta^*)\hat{\lambda}$ is asymptotically negligible. By Lemma \ref{lambdapro} and Lemma D.2 in  Kitamura et al. (2004),  we have $\|\hat{\lambda}\|=O_p(n^{-1/2})$ and $\max_{1\leq i\leq n}\{\|g(\delta_i,X_i,Z_i,Y_i, \theta^*)\|\}=o_p(n^{1/2})$. Then,
\begin{eqnarray*}\label{A22}
&&\left|\sum_{i=1}^nA_{2i}(\eta)\delta_ig^T(\delta_i,X_i,Z_i,Y_i, \theta^*)\hat{\lambda}\right|\nonumber\\
&\leq& \max_{1\leq i\leq n}\{\|g(\delta_i,X_i,Z_i,Y_i, \theta^*)\|\}\|\hat{\lambda}\|\Bigr| \sum_{i=1}^n \delta_i A_{2i}(\eta)\Bigr|=o_p(1).
\end{eqnarray*}

By the asymptotic expressions of (\ref{AA1}) and (\ref{AA2}), we conclude that $A(\eta)\drow A_0(\eta)$, where
\begin{eqnarray*}
A_0(\eta)=\eta^Tn^{-1/2}\left\{  U_\phi(\theta^*)-F_gS_g^{-1} \left[U_g(\theta^*) +  G_\gamma S_B^{-1}  U(\gamma^*)\right]\right\}+\frac{1}{2}\eta^T F_\beta\eta.
\end{eqnarray*}
Then, it follows that
\begin{eqnarray*}
\sqrt{n}(\hat{\beta}_{ELW}-\beta^*)=\hat{\eta}\drow \arg\min_{\eta} A_0(\eta),
\end{eqnarray*}
where
\begin{eqnarray*}
\arg\min_{\eta} A_0(\eta)&=&-F_\beta^{-1}n^{-1/2}\left\{  U_\phi(\theta^*)-F_gS_g^{-1} \left[U_g(\theta^*) +  G_\gamma S_B^{-1}  U(\gamma^*)\right]\right\}.
\end{eqnarray*}
Furthermore, by simple algebra, one can verify that
\begin{eqnarray*}
 \left(\begin{array}{cc}
                  D_3 & D_4
                \end{array}
\right)\left(\begin{array}{cc}
 D_1& D_2 \\
 D_2^T & S_B                                                                                          \end{array}\right)^{-1}&=&  \left(
                                    \begin{array}{cc}
                                       V_1V_2^{-1} &  D_4S_B^{-1}-V_1V_2^{-1}D_2S_B^{-1} \\
                                    \end{array}
                                  \right)
\end{eqnarray*}
and
\begin{eqnarray*}
&&U_g(\theta^*) +  G_\gamma S_B^{-1}  U(\gamma^*)\\
&=& \left(\begin{array}{c}\sum_{i=1}^n [g_1(\delta_i,X_i,Z_i,Y_i, \theta^*) -D_2 S_B^{-1} U_{B}(\delta_i,Z_i,Y_i,\gamma^*)] \\ 0 \end{array} \right).
\end{eqnarray*}
Therefore,
\begin{eqnarray*}
 &&F_gS_g^{-1} \left[U_g(\theta^*) +  G_\gamma S_B^{-1}  U(\gamma^*)\right]\\
 &=& \left(\begin{array}{cc}
                  D_3 & D_4
                \end{array}
\right)\left(\begin{array}{cc}
D_1& D_2 \\
D_2^T & S_B  \end{array}\right)^{-1}
\left[U_g(\theta^*) +  G_\gamma S_B^{-1}  U(\gamma^*)\right]\\
&=&   V_1V_2^{-1} \sum_{i=1}^n\left[g_1(\delta_i,X_i,Z_i,Y_i, \theta^*) -D_2 S_B^{-1} U_{B}(\delta_i,Z_i,Y_i,\gamma^*) \right].
\end{eqnarray*}
Let
\begin{eqnarray*}
    h_{1i}&=&   \delta_i\phi(X_i,Z_i,Y_i,\beta^*),\\
    h_{2i}&=&  g_1(\delta_i,X_i,Z_i,Y_i, \theta^*) -D_2 S_B^{-1} U_{B}(\delta_i,Z_i,Y_i,\gamma^*).
\end{eqnarray*}
 One can write $\arg\min_{\eta} A_0(\eta)$ as
$-F_\beta^{-1}n^{-1/2}  \sum_{i=1}^n  \left\{ h_{1i}-V_1V_2^{-1} h_{2i}\right\}$.
It is easily verified that $Var\left(h_{1i}\right)= E(h_{1i}h_{1i}^T)=S_\phi,$
\begin{eqnarray*}
  E (h_{1i}h_{2i}^T)&=&E(h_{1i} g_1^T(\delta_i,X_i,Z_i,Y_i, \theta^*))-   E(h_{1i} U_{B}^T(\delta_i,Z_i,Y_i,\gamma^*))S_B^{-1}D_2^T \\
                    &=&D_3-D_4S_B^{-1}D_2^T,
\end{eqnarray*}
and
\begin{eqnarray*}
 Var\left(h_{2i}\right)&=& Var(g_1(\delta_i,X_i,Z_i,Y_i, \theta^*)) + D_2 S_B^{-1} Var(U_{B}(\delta_i,Z_i,Y_i,\gamma^*))S_B^{-1}D_2^T  \\
 &&-E[ g_1(\delta_i,X_i,Z_i,Y_i, \theta^*)U_{B}^T(\delta_i,Z_i,Y_i,\gamma^*)] S_B^{-1}D_2^T\\
 &&- D_2 S_B^{-1} E[U_{B}(\delta_i,Z_i,Y_i,\gamma^*)g_1^T(\delta_i,X_i,Z_i,Y_i, \theta^*) ] \\
 &=&D_1-D_2 S_B^{-1}D_2^T.
\end{eqnarray*}
Thus,
\begin{eqnarray*}
&&Var\left(   h_{1i}-V_1V_2^{-1} h_{2i} \right)\\
&=& Var ( h_{1i} )-V_1V_2^{-1} E ( h_{2i}h_{1i}^T) -E( h_{1i}h_{2i}^T)V_2^{-1} V_1^T+ V_1V_2^{-1}Var(h_{2i})V_2^{-1} V_1^T\\
&=& S_\phi-V_1V_2^{-1}(D_3-D_4 S_B^{-1}D_2^T )^T-  (D_3-D_4 S_B^{-1}D_2^T )V_2^{-1} V_1^T\\
&&+V_1V_2^{-1}( D_1-D_2 S_B^{-1}D_2^T  )V_2^{-1} V_1^T \\
&=& S_\phi-V_1V_2^{-1} V_1^T.
\end{eqnarray*}
The desired result follows by the central limit theorem.
\\
\textbf{\emph{ The proof of Theorem \ref{el} }}
According to the proof of  Theorem 1 of Lopez et al.(2009),
it can be shown that
\begin{eqnarray*}
n^{1/2}\left(\begin{array}{c}
\hat{\beta}_{EL}-\beta^*\\
\hat{\gamma}_{EL}-\gamma^*\\
\end{array}\right)
\drow N\left(0, \Sigma_{EL} \right),
\end{eqnarray*}
 where $\Sigma_{EL}=\left(S_1^T S_2^{-1}S_1\right)^{-1}$,
$
S_1=\left(\begin{array}{cc}
                                                                                                          F_\beta & 0 \\
                                                                                                         0 & D_2 \\
                                                                                                         0 & S_B
                                                                                                       \end{array}
\right)
$
and
$
S_2=\left(\begin{array}{ccc}
                                                                                                          S_\phi &D_3 &D_4 \\
                                                                                                         D_3^T & D_1&D_2 \\
                                                                                                         D_4^T &D_2^T & S_B
                                                                                                       \end{array}
\right).
$

Let $E_{22}=\left(\begin{array}{cc} D_1& D_2 \\ D_2^T & S_B \end{array}\right)$, then we write
$S_2=\left(\begin{array}{cc} S_\phi & F_g \\ F_g^T & E_{22} \end{array}\right)$, where $F_g$ is defined in (\ref{Fg})
We know  that
 \begin{eqnarray*}
S_2^{-1}=\left( \begin{array}{cc}
                  E_{11.2}^{-1} & -E_{11.2}^{-1}F_gE_{22}^{-1}   \\
                  -E_{22}^{-1}F_g^TE_{11.2}^{-1}  & E_{22}^{-1}+E_{22}^{-1} F_g^TE_{11.2}^{-1}F_gE_{22}^{-1}
                \end{array}
 \right),
\end{eqnarray*}
 with $ E_{11.2}= S_\phi-F_gE_{22}^{-1}F_g^T=S_\phi-V_1V_2^{-1}V_1^T-D_4S_B^{-1}D_4^T$.
Note that $S_1^T S_2^{-1}S_1$ can  be written as
\begin{eqnarray*}
 &&S_1^T S_2^{-1}S_1  \\
  &=&  \left(\begin{array}{cc}
                            F_\beta  E_{11.2}^{-1} F_\beta  & -F_\beta E_{11.2}^{-1}F_gE_{22}^{-1}\left(\begin{array}{c}D_2\\S_B \end{array}\right)
                              \\
                           -  (D_2^T \ S_B) E_{22}^{-1}F_g^TE_{11.2}^{-1}F_\beta   &   (D_2^T \ S_B)\{E_{22}^{-1}+E_{22}^{-1} F_g^TE_{11.2}^{-1}F_gE_{22}^{-1} \} \left(\begin{array}{c}D_2\\S_B \end{array}\right)
                          \end{array}\right)
   \\
    &=&  \left(\begin{array}{cc}
                            F_\beta  E_{11.2}^{-1} F_\beta  & -F_\beta E_{11.2}^{-1}D_4
                              \\
                           - D_4^TE_{11.2}^{-1}F_\beta   &   S_B+ D_4^T E_{11.2}^{-1} D_4
                          \end{array}\right) = \left(\begin{array}{cc}
                           H_{11}  &   H_{12}
                              \\
                            H_{21}   &   H_{22}
                          \end{array}\right),
\end{eqnarray*}
where $ H_{11}= F_\beta  E_{11.2}^{-1}F_\beta  $, $ H_{12}=-F_\beta E_{11.2}^{-1}D_4$, $H_{21}=H_{12}^T$ and $H_{22}=S_B+ D_4^T E_{11.2}^{-1} D_4$.
Therefore, we have
\begin{eqnarray*}
 (S_1^T S_2^{-1}S_1)^{-1}
    &=&   \left(\begin{array}{cc}
                            H_{11}  &   H_{12}
                              \\
                            H_{21}   &   H_{22}
                          \end{array}\right)^{-1}\\
    &=&\left(\begin{array}{cc}
                           H_{11}^{-1}+H_{11}^{-1}H_{12} H_{22.1}^{-1}H_{21}H_{11}^{-1}    & - H_{11}^{-1}H_{12} H_{22.1}^{-1}
                              \\
                           - H_{22.1}^{-1}H_{21}H_{11}^{-1}  &    H_{22.1}^{-1}
                          \end{array}\right),
\end{eqnarray*}
where $ H_{22.1}=H_{22}-H_{21}H_{11}^{-1}H_{12}=S_B.$
By direct calculation, it follows that
\begin{eqnarray*}
  &&H_{11}^{-1}+H_{11}^{-1}H_{12} H_{22.1}^{-1}H_{21}H_{11}^{-1}\\
   &=& F_\beta^{-1} E_{11.2}F_\beta^{-1}+F_\beta^{-1} D_4 S_B^{-1}  D_4^T F_\beta^{-1} \\
  &=& F_\beta^{-1}S_\phi F_\beta^{-1}-F_\beta^{-1}V_1V_2^{-1}V_1^TF_\beta^{-1}-F_\beta^{-1}D_4S_B^{-1}D_4^TF_\beta^{-1}+F_\beta^{-1} D_4 S_B^{-1}  D_4^T F_\beta^{-1}\\
  &=& F_\beta^{-1}S_\phi F_\beta^{-1}-F_\beta^{-1}V_1V_2^{-1}V_1^TF_\beta^{-1}\\
  &=&\Sigma_{ELW},
\end{eqnarray*}
and
$
-H_{11}^{-1}H_{12} H_{22.1}^{-1}  = F_\beta^{-1} E_{11.2}F_\beta^{-1} F_\beta E_{11.2}^{-1}D_4 S_B^{-1}
  =F_\beta^{-1}  D_4 S_B^{-1} .
$
The desired result follows.


\section*{Reference}

\begin{thebibliography}{00}




\bibitem[(Bartlett2014)]{Bartlett2014}
Bartlett J W, Carpenter J R, Tilling K, et al. 2014. Improving upon the efficiency of complete case analysis when covariates are MNAR[J]. Biostatistics, 15(4): 719-730.

\bibitem[(Chen2015)]{Chen2015}
Chen X, Wan A T K, Zhou Y. 2015. Efficient quantile regression analysis with missing observations[J]. Journal of the American Statistical Association, 110(510):00-00.

\bibitem[(Kitamura)]{Kitamura}
Kitamura Y, Tripathi G, Ahn H. 2004. Empirical likelihood-based inference in conditional moment restriction Models[J]. Econometrica, 72(6): 1667-1714.
\bibitem[(Knight1998)]{Knight1998}
Knight K. 1998. Limiting distributions for $L_1$ regression estimators under general conditions[J]. Annals of Statistics, 26: 755-770.

\bibitem[(Koenker1978)]{Koenker1978}
Koenker, R. and Bassett, G. 1978. Regression quantiles. Econometrica,46, 33-50.

\bibitem[(Koenker2005)]{Koenker2005}
Koenker R. 2005. Quantile regression[M]. Cambridge university press.

\bibitem[(Little2002)]{Little2002}
Little RJA,  Rubin DB 2002  Statistical analysis with missing data, 2nd ed, Wiley, Hoboken, NJ.

\bibitem[(Little2011)]{Little2011}
Little, R.J., Zhang, N. 2011. Subsample ignorable likelihood for regression analysis with missing data. \emph{J R Stat Soc Ser C} 60:
591-605.



\bibitem[(Liu2012)]{Liu2012}
Liu, T. and Yuan, X. 2012. Combining quasi and empirical likelihoods in generalized linear models
with missing responses. Journal of Multivariate Analysis   111, 39-58.

\bibitem[(Liu2013)]{Liu2013}
Liu, T., Yuan, X., Li, Z. and Li, Y. 2013. Empirical and weighted conditional likelihoods for matched
case-control studies with missing covariates. Journal of Multivariate Analysis  119, 185-199.

\bibitem[(Liu2016)]{Liu2016}
Liu T, Yuan X. 2016. Weighted quantile regression with missing covariates using empirical likelihood[J]. Statistics A Journal of Theoretical and Applied Statistics, 50(1):89-113.

\bibitem[(Liu2016)]{Lopez2009}
Lopez  EMM, Keilegom IV, Veraverbeke N (2009) Empirical likelihood for non-smooth criterion functions.  Scand J Stat 36: 413-432.


\bibitem[(Owen01)]{Owen01}
Owen, A. B. 2001. Empirical Likelihood. Chapman and Hall, New York.

\bibitem[(Owen01)]{Qin1994}
 Qin J, Lawless J 1994 Empirical likelihood and general estimating equations. Ann Stat  22(1): 300-325.


\bibitem[(Qin)]{Qin2009}
Qin, J., Zhang, B.  and Leung, D. H. 2009. Empirical likelihood in missing data problems[J]. Journal of the American Statistical Association,  104:
1492-1503.



\bibitem[(Sherwood2013)]{Sherwood2013}
Sherwood B, Wang L, Zhou X H. 2013. Weighted quantile regression for analyzing health care cost data with missing covariates.[J]. Statistics in Medicine, 32(28):4967-79.

\bibitem[(wang)]{Wang02}
Wang, Q., and Rao, J. N. K. 2002. Empirical likelihood-based inference
under imputation for missing response data[J]. Annals of Statistics, 30: 896-924.

\bibitem[(Wei2012)]{Wei2012}
Wei Y, Ma Y, Carroll R J.2012. Multiple imputation in quantile regression[J]. Biometrika,  99(2):423-438.

 \bibitem[(Wei2012)]{Xie2017}
Xie, Y., Zhang, H. 2017. Empirical likelihood in nonignorable covariate-missing data problems.\emph{ Int. J. Biostat} 13(1): 20160053.
 

\bibitem[(Wei2012)]{Zhong2017}
Zhong G,  Qin J. 2017. Empirical likelihood method for non-ignorable missing data problems. Lifetime Data Anal  23(1):113-135.



\end{thebibliography}


\begin{table}\begin{center}
Table 1: Empirical bias and RMSE (in parentheses)  based on 1000
simulations with $n = 100,300$.\\
\label{tab1}       
\begin{tabular}{ccc ccc }
 \hline
\multicolumn{1}{c}{$\tau$}&\multicolumn{1}{c}{n}&Estimator&$\beta_0$ &$\beta_1$ & $\beta_2$ \\
\hline

0.3&100&$\hat{\beta}_{ideal}$&-0.0022 (0.1185)&0.0084 (0.1095)&-0.0062 (0.1272)  \\
   &   &$\hat{\beta}_C$       &0.0072 (0.2403)&0.0032 (0.1851)&-0.0030 (0.1853)  \\
   &   &$\hat{\beta}_{IPWMAR}$& -0.1714(0.3065)&0.0130 (0.2128)&-0.0042 (0.2077)  \\
   &   &$\hat{\beta}_{ELWMAR}$&-0.1808 (0.3021)&0.0231 (0.2144)&-0.0120 (0.1945)  \\
   &   &$\hat{\beta}_{ELW}$  &-0.0004 (0.2446)&0.0107 (0.1875)&-0.0098 (0.1752)  \\

   &300&$\hat{\beta}_{ideal}$ &0.0016 (0.0685)&0.0001 (0.0617)&-0.0014 (0.0676)  \\
   &   &$\hat{\beta}_C$       &0.0008 (0.1332)&0.0032 (0.1031)&-0.0011 (0.1016)  \\
   &   &$\hat{\beta}_{IPWMAR}$&-0.1669 (0.2188)&0.0133 (0.1167)&-0.0068 (0.1150)  \\
   &   &$\hat{\beta}_{ELWMAR}$&-0.1672 (0.2079)&0.0155 (0.1116)& -0.0097(0.0968)  \\
   &   &$\hat{\beta}_{ELW}$  &-0.0007 (0.1252)&0.0053 (0.1002)&-0.0032 (0.0873)  \\\\

0.5&100&$\hat{\beta}_{ideal}$ &-0.0021 (0.1128)&0.0018 (0.0997)&0.0038 (0.1187)  \\
   &   &$\hat{\beta}_C$       &0.0016 (0.2347)&0.0014 (0.1781)&0.0073 (0.1765)  \\
   &   &$\hat{\beta}_{IPWMAR}$&-0.1636 (0.2761)&0.0148 (0.1851)&-0.0020 (0.1850)  \\
   &   &$\hat{\beta}_{ELWMAR}$&-0.1693 (0.2794)&0.0209 (0.1814)&-0.0076 (0.1649)  \\
   &   &$\hat{\beta}_{ELW}$  &-0.0023 (0.2326)&0.0042 (0.1685)&0.0007 (0.1617)  \\

   &300&$\hat{\beta}_{ideal}$ &-0.0015 (0.0648)&0.0002 (0.0608)&0.0011 (0.0679)  \\
   &   &$\hat{\beta}_C$       &-0.0001 (0.1274)&0.0008 (0.0979)&0.0017 (0.0973)  \\
   &   &$\hat{\beta}_{IPWMAR}$&-0.1646 (0.2040)&0.0136 (0.1036)&-0.0073 (0.1003)  \\
   &   &$\hat{\beta}_{ELWMAR}$&-0.1650 (0.2007)&0.0158 (0.1002)&-0.0087 (0.0891)  \\
   &   &$\hat{\beta}_{ELW}$  &-0.0017 (0.1238)&0.0032 (0.0954)&-0.0001 (0.0889)  \\\\

0.7&100&$\hat{\beta}_{ideal}$ &-0.0090 (0.1216)&0.0025 (0.1147)&0.0021 (0.1212)  \\
   &   &$\hat{\beta}_C$       &-0.0232 (0.2471)&0.0118 (0.1864)&-0.0042 (0.1791)  \\
   &   &$\hat{\beta}_{IPWMAR}$&-0.1750 (0.2839)&0.0253 (0.1845)&-0.0089 (0.1790)  \\
   &   &$\hat{\beta}_{ELWMAR}$&-0.1772 (0.2901)&0.0263 (0.1868)&-0.0105 (0.1694)  \\
   &   &$\hat{\beta}_{ELW}$  &-0.0203 (0.2498)&0.0076 (0.1870)&0.0002 (0.1795) \\

   &300&$\hat{\beta}_{ideal}$ &-0.0014 (0.0706)&0.0019 (0.0630)&0.0001 (0.0708)  \\
   &   &$\hat{\beta}_C$       &0.0018 (0.1371)&-0.0032 (0.1008)&0.0019 (0.1028)  \\
   &   &$\hat{\beta}_{IPWMAR}$&-0.1576 (0.2007)&0.0136 (0.1046)&-0.0062 (0.1033)  \\
   &   &$\hat{\beta}_{ELWMAR}$&-0.1557 (0.1983)&0.0130 (0.1006)&-0.0059 (0.0935)  \\
   &   &$\hat{\beta}_{ELW}$  &0.0025 (0.1325)&-0.0003 (0.0969)&0.0009 (0.0964)  \\

\hline
\end{tabular}\end{center}
\end{table}

\newpage
\begin{figure}
\centering
 \includegraphics[width=1.0\textwidth, angle=
0]{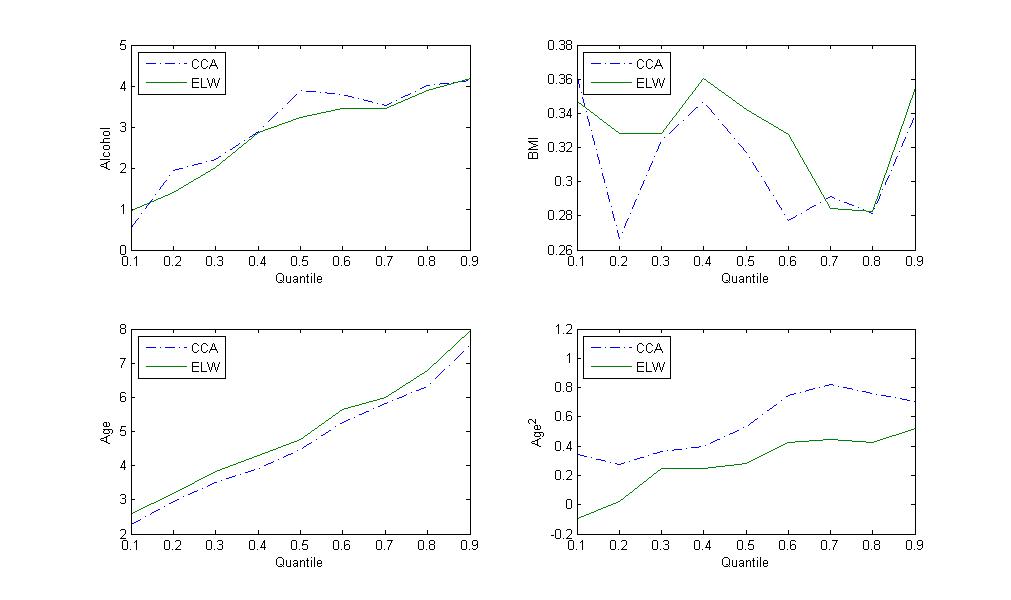}   {Figure 1: The estimated regression coefficients, $\hat{\beta}_C$ (-.) and $\hat{\beta}_{ELW}$ (--) at various quantile levels.}\label{fig1}
\end{figure}
\newpage
\begin{center}
\begin{figure}
\centerline{\includegraphics[width=1.0\textwidth, angle=
0]{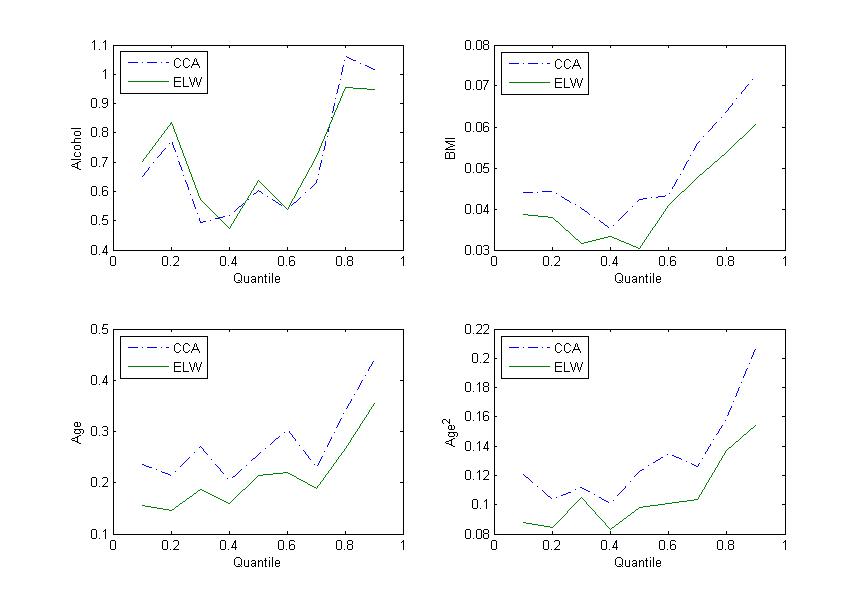}}  {Figure 2: The standard errors of $\hat{\beta}_C$ (-.) and $\hat{\beta}_{ELW}$ (--) at various quantile levels.}\label{fig2}
\end{figure}
\end{center}

\end{document}